\documentstyle[multicol,aps,prl,epsf]{revtex}

\begin{document}

\draft

\title{
A consistent description of the pairing symmetry in hole and electron
doped cuprates within the two dimensional Hubbard model
}

\author{
Kazuhiko Kuroki and Hideo Aoki
}
\address{Department of Physics, University of Tokyo, Hongo,
Tokyo 113, Japan}

\date{\today}

\maketitle

\begin{abstract}
Quantum Monte Carlo is used to calculate various pairing 
correlations of the 2D Hubbard model possessing
band features experimentally observed in the cuprates.
In the hole-doped case, where the Fermi level lies 
close to the van Hove singularities around $(0,\pi)$,
the d$_{x^2-y^2}$ pairing correlation is selectively enhanced, while 
in the electron-doped case, where 
the singularities are far below the Fermi level and 
the Fermi surface runs through $(\pm \pi/2,\pm \pi/2)$, 
both d$_{x^2-y^2}$ and d$_{xy}$ correlations are enhanced with the latter
having a $\sqrt{2}\times \sqrt{2}$ structure.  
The two pairing symmetries
can mix to result in a nodeless gap.
\end{abstract}

\medskip


\begin{multicols}{2}
\narrowtext

\newpage
Since the seminal proposal by Anderson,\cite{Anderson} great theoretical 
effort has been made to investigate the possibility of describing 
various aspects of the high T$_C$ cuprates within the 
two dimensional (2D) Hubbard model.\cite{Dagrev} Those include the
antiferromagnetic insulating phase in the undoped systems,
the normal state above T$_C$, and the superconducting state.
Among all, it has been an open question whether the Hubbard model
can actually account for the superconductivity, especially its
pairing symmetry, in both hole and electron doped cuprates.

Many analytical calculations have supported the possibility of 
d$_{x^2-y^2}$ pairing in the nearly half-filled 2D Hubbard model.
\cite{Bickers,Dzyalo,Schulz,Alvarez}  
While some previous numerical studies of pairing correlations 
in finite systems have given negative results for superconductivity,
\cite{Furukawa,Moreo,Zhang}  
we have recently shown that an enhanced 
d$_{x^2-y^2}$ pairing correlation is indeed detected numerically 
if we ensure that the highest occupied one-electron levels 
(HOL) and lowest unoccupied levels (LUL) at $U=0$ in finite systems 
are sufficiently close.\cite{KA}  
This precaution, as motivated from the numerical studies on Hubbard ladders,
\cite{Yamaji,KTA} has been necessitated because 
the energy scale of the superconductivity in the Hubbard 
model, if any, should be of the order of 
$0.01t$,\cite{Bickers,Alvarez} while the discreteness of the energy levels 
in finite systems tractable in numerical calculation 
is much larger ($\sim 0.1t$) unless parameter values are tuned. 

In the hole-doped cuprates such as YBCO and BSCCO 
there is now a body of accumulating evidence that the pairing symmetry is 
d$_{x^2-y^2}$ 
(at least around the optimal doping),\cite{Scalrev} which is consistent
with the previous Hubbard-model studies.  
On the other hand, experimental results for the  
electron-doped NCCO seems to indicate an s-wave, or more precisely, 
a symmetry with a nodeless superconducting gap. 
\cite{Wu,Andreone,Anlage,Stadlober,Kashiwaya}
Experiments have also revealed further 
differences between hole-doped and electron-doped systems. 
Specifically, the angle-resolved photoemission spectroscopy (ARPES)
has shown that the `extended' van Hove singularity (VHS) around
${\bf k}=(0,\pi)$ and $(\pi,0)$ lies 
very close to the Fermi level $(E_F)$ in YBCO and BSCCO, while the VHS lies
far below the Fermi level in NCCO.\cite{Shen} 

The purpose of the present study is to explore whether 
the difference of the pairing symmetry between 
electron and hole doped systems can be explained within the 
2D Hubbard model possessing the band features observed experimentally.
The essential band features (namely, the shape of the Fermi surface 
and the relative position of the VHS to $E_F$) of YBCO, BSCCO, and NCCO 
can be reproduced by introducing a next-nearest neighbor (NNN)
transfer about half the nearest neighbor (NN) one.
Note that in our previous study mentioned above\cite{KA}, 
such an electron-hole asymmetry was not taken into account since
we considered the Hubbard model with only NN transfers.

Quantum Monte Carlo (QMC) method is used to calculate 
correlation functions of d$_{xy}$, NN and NNN extended s 
as well as d$_{x^2-y^2}$ pairings.  
To look into such diverse symmetries has been motivated from 
the following physical consideration. Namely, the pair-tunneling 
processes between $({\bf k}_1\uparrow, -{\bf k}_1\downarrow)$ and 
$({\bf k}_2\uparrow, -{\bf k}_2\downarrow)$ favors the pairing order
parameter $\Delta$ that satisfies $\Delta({\bf k}_1)=-\Delta({\bf k}_2)$, 
which is a picture known to be at work in the two-leg
\cite{Fabrizio,Schulz2,Balents} 
and three-leg\cite{Schulz2,Arrigoni,TKA,LBF} Hubbard ladders. 
In this picture,
the pair-tunneling between the $k$-points around $(0,\pi)$ 
and $(\pi,0)$ favors d$_{x^2-y^2}$ pairing.  
Such processes should indeed be pronounced in the hole-doped cuprates
because $E_F$ lies close to $(0,\pi)$ and $(\pi,0)$, and the density
of states around these points is large.
Thus, in this case, d$_{x^2-y^2}$ pairing 
should be dominant with possibly other symmetries mixing slightly. 
By contrast, in the
electron doped case, other pair-tunneling processes may
set in on a nearly equal footing 
in determining the pairing symmetry, because
VHS lies far below $E_F$. Then, not only d$_{x^2-y^2}$ but also
d$_{xy}$ pairing or extended s pairing 
(with gap functions that have nodes on the Fermi surface, 
but do not change sign by a 90 degree rotation) 
will become eligible, so that 
some of these symmetries may mix with comparable weights.   

In fact, we find here 
that the d$_{x^2-y^2}$ correlation is dominant 
in the hole-doped case, while in the electron-doped case 
both d$_{x^2-y^2}$ and d$_{xy}$ correlations are enhanced, with
the latter having a $\sqrt{2}\times\sqrt{2}$ structure. 
These two pairing symmetries can in fact mix ending up with a nodeless
gap without breaking the time reversal symmetry, 
unlike in d$_{x^2-y^2}+i$d$_{xy}$ pairing\cite{Rokhsar,Laughlin} 
where the symmetry is broken. 
Correlation of the extended s-wave pairings is 
found to be suppressed in all the cases investigated.

We consider the 2D Hubbard model on a square lattice with NN ($t$), 
NNN ($t'$), and third NN ($t''$) hoppings,
\begin{eqnarray*}
{\cal H}&=&-\sum_{x,y,\sigma}
\left[t_x(c_{x,y,\sigma}^\dagger c_{x+1,y,\sigma})
+t_y(c_{x,y,\sigma}^\dagger c_{x,y+1,\sigma})\right.\nonumber\\
&+&t'_{/}(c_{x,y,\sigma}^\dagger c_{x+1,y+1,\sigma})
+t'_{\backslash}(c_{x,y,\sigma}^\dagger c_{x-1,y+1,\sigma})\nonumber\\
&+&\left. t''(c_{x,y,\sigma}^\dagger c_{x+2,y,\sigma}
+c_{x,y,\sigma}^\dagger c_{x,y+2,\sigma})+{\rm h.c.}\right]\nonumber\\
&+&U\sum_{x,y} n_{x,y,\uparrow}n_{x,y,\downarrow}.
\end{eqnarray*}
Here, $(x,y)$ is the coordinate of the sites, and 
the lattice constant is taken as unity. 
Periodic boundary condition is assumed, and we set $t_x=1$ hereafter.

As mentioned above\onlinecite{KA},
it is necessary to put $E_F$ at $U=0$ between the 
HOL's and LUL's separated by an energy of 
$\Delta\varepsilon^0$ less than $O(0.01)$ in order to
detect a symptom of superconductivity having an energy 
scale of $O(100K)$.  
On the other hand, QMC is unstable 
for exactly $\Delta\varepsilon^0=0$, namely for open shell configurations.  
Thus, we accomplish $\Delta\varepsilon^0 \sim O(0.01)$
by making $t_x$ and $t_y$, and/or $t'_{/}$ and $t'_{\backslash}$ 
slightly different, where 
$t_x\neq t_y$ lifts the degeneracy between $(\pm k_1,\pm k_2)$ and 
$(\pm k_2,\pm k_1)$ for $|k_1|\neq |k_2|$, 
while $t'_{/}\neq t'_{\backslash}$ 
lifts the degeneracy between $(k_1,k_1)$ and 
$(\pm k_1,\mp k_1)$.  

We have employed the ground-state, 
canonical-ensemble QMC, where we have implemented
the stabilization algorithm adopted by several
authors.\cite{stab} We adopt the free 
Fermi sea as the trial state, and take the projection 
imaginary time $\tau$ up to $\sim 40$ to ensure the convergence. 
Small $\Delta\varepsilon^0$ makes the negative sign problem serious,
but by taking a relatively small value of $U(=1)$, 
we can check the convergence with respect to $\tau$ 
without running into serious sign problem.

We have calculated the pairing correlation functions,
\begin{eqnarray*}
&&P(r)=\sum_{|\Delta x|+|\Delta y|=r} 
\langle O^\dagger (x+\Delta x,y+\Delta y) O (x,y)\rangle\:\:\:
{\rm with}\nonumber\\
&&O_{\rm NN}(x,y)=\sum_{\delta,\sigma}
\sigma(c_{x,y,\sigma} c_{x+\delta,y,-\sigma}
\pm c_{x,y,\sigma} c_{x,y+\delta,-\sigma})\nonumber\\
&&O_{\rm NNN}(x,y)=\sum_{\delta,\sigma}
\sigma(c_{x,y,\sigma} c_{x+\delta,y+\delta,-\sigma}
\pm c_{x,y,\sigma} c_{x-\delta,y+\delta,-\sigma}),
\end{eqnarray*}
where $\delta=\pm 1$.  
The plus (minus) sign in $O_{\rm NN}$ corresponds to NN s 
(d$_{x^2-y^2}$) symmetries, while 
the plus (minus) in $O_{\rm NNN}$ to 
NNN s (d$_{xy}$) symmetries.  
\begin{figure}
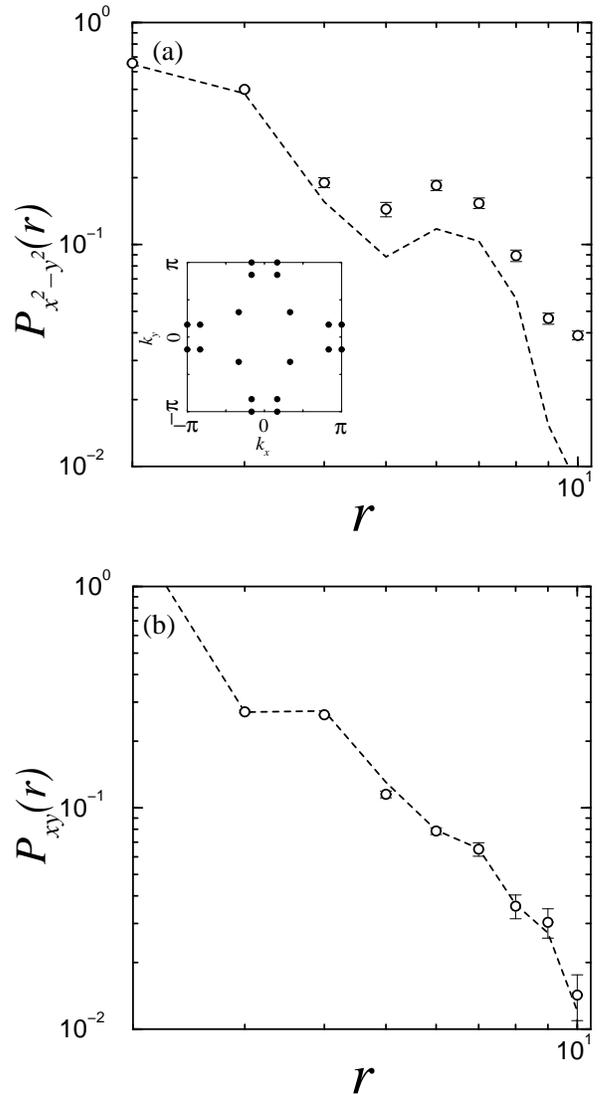

\begin{center}
\leavevmode\epsfysize=70mm \epsfbox{condfig1a.eps}
\end{center}
\begin{center}
\leavevmode\epsfysize=70mm \epsfbox{condfig1b.eps}
\caption{
QMC result for d$_{x^2-y^2}$ (a) and d$_{xy}$ (b) pairing correlations 
for a hole-doped system ($12\times 12$ system with 118 electrons 
with $n=0.82$).
$t_y=0.999$, $t'_{/}=-0.429$, 
$t'_{\backslash}=-0.43$, 
$t''=0.07$, and $U=1$ ($\bigcirc$).
The dashed lines represent the $U=0$ result. 
The inset shows the HOL's and LUL's within
0.01 to $E_F$. 
}
\label{fig1}
\end{center}
\end{figure}

We have looked into various values of $n$, $t'$, and $t''$ 
including other than the ones described below, and found 
NN and NNN s-wave pairing correlations to be 
strongly suppressed with $U$ at large distances. 
At first this may seem odd 
because these pairings do not
have any on-site amplitude.
This might be because the extended s pairings 
always couple, at least at the mean-field level, 
with the on-site s pairing,\cite{Micnas} 
which is directly suppressed with $U>0$.
Thus, we show only d$_{x^2-y^2}$ and d$_{xy}$
pairing correlations in the following. 

We first look at the 
hole-doped case. We consider a $12\times 12$ lattice with 
118 electrons (band filling $n=0.82)$ with $t_y=0.999$, $t'_{/}=-0.429$, 
$t'_{\backslash}=-0.43$, and $t''=0.07$.  
For this choice, 
the HOL's\cite{comment} at $(\pm\pi/6,\pi)$, $(\pm\pi/6,\pm5\pi/6)$, 
$(\pi/3,-\pi/3)$, $(-\pi/3,\pi/3)$
and LUL's at 
$(\pi,\pm\pi/6)$, $(\pm5\pi/6,\pm\pi/6)$, 
$(\pi/3,\pi/3)$, $(-\pi/3,-\pi/3)$ 
lie within 0.01 in energy at $U=0$.  
The Fermi surface, represented by these HOL's and LUL's is displayed
as an inset in Fig.\ref{fig1}(a).
There, reflecting the high density of states around VHS, 
many $k$-points around $(0,\pi)$ and $(\pi,0)$ appear, 
while the points around $|k_x|=|k_y|$, although fewer, also exist. 

In Fig.\ref{fig1}(a), we show the d$_{x^2-y^2}$ correlation
as a function of real space distance $r\equiv |\Delta x|+ |\Delta y|$. 
It can be seen that
the correlation is enhanced for $U=1$ over that for $U=0$,
especially at large distances. By contrast, the d$_{xy}$
correlation shown in (b) is not enhanced within the error bars. 
The dominant d$_{x^2-y^2}$ pairing is consistent with the 
expectation from the pair-tunneling picture given above.  
On the other hand, we cannot rule out 
the possibility of a small d$_{xy}$ mixing, since 
if more $k$-points exist in the vicinity of $E_F$, not only 
the d$_{x^2-y^2}$ correlation would be more enhanced,
but also the d$_{xy}$ might be enhanced, which would imply 
their mixture.  
Further, d$_{xy}$ may mix in a
time-reversal broken form, 
d$_{x^2-y^2}$+id$_{xy}$,\cite{Rokhsar,Laughlin}
especially in magnetic fields,\cite{Laughlin2} which 
is of interest from the viewpoint of the recent experimental observations
suggesting such a possibility at low temperatures.\cite{Krishana,Movshovich}

\begin{figure}
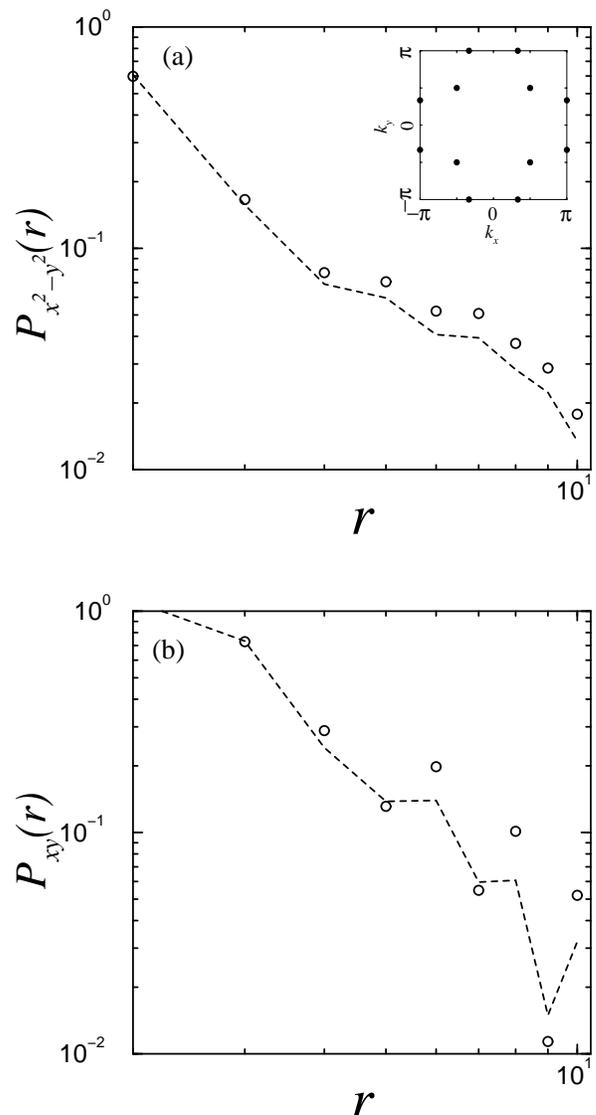

\begin{center}
\leavevmode\epsfysize=70mm \epsfbox{condfig2a.eps}
\end{center}
\begin{center}
\leavevmode\epsfysize=70mm \epsfbox{condfig2b.eps}
\caption{
A plot similar to Fig.\protect\ref{fig1} for 
an electron-doped system 
($190$ electrons $/12\times 12$ with $n=1.32$) for 
$t_y=0.999$, $t'_{/}=-0.499$, 
$t'_{\backslash}=-0.5$, 
$t''=0$, and $U=1$.
}
\label{fig2}
\end{center}
\end{figure}

Let us now turn to the case of electron doping. This time, we take 
$190$ electrons $/ 12\times 12$ $(n=1.32)$ with $t_y=0.999$, 
$t'_{/}=-0.499$, $t'_{\backslash}=-0.5$, and $t''=0$. 
(In the actual calculation
we have employed the electron-hole transformation to 
consider a 98 electron system with $t'>0$). 
Here, HOL's reside at $(\pm\pi/3,\pi)$, $(\pi/2,-\pi/2)$, $(-\pi/2,\pi/2)$,
while LUL's at $(\pi,\pm\pi/3)$, $(\pi/2,\pi/2)$, $(-\pi/2,-\pi/2)$
for $U=0$ (inset of Fig.\ref{fig2}(a)). 
Note that $(\pm\pi/2,\pm\pi/2)$
lies right on the Fermi surface, a feature seen in the ARPES data
of NCCO.\cite{King} 

The QMC result in Fig.\ref{fig2}(a) shows that, although the Fermi surface is 
now shifted away from $(\pi,0)$, $(0,\pi)$ down to 
$(\pi,\pm\pi/3)$, $(\pm\pi/3,\pi)$, 
we still have an enhancement of the d$_{x^2-y^2}$ correlation, 
although the enhancement is smaller
than that in the holed-doped case. 

Now, more striking is the behavior of the d$_{xy}$ correlation 
shown in Fig.2(b). 
At large distances, the d$_{xy}$ correlation is enhanced at even distances 
($\Delta x+ \Delta y$ = even), 
while suppressed at odd distances, which means that it has a 
$\sqrt{2}\times \sqrt{2}$ {\it superstructure}. A Fourier transform 
of the correlation function indeed shows that its $(\pi,\pi)$ component 
is enhanced with $U$.

The result suggests a coexistence of the d$_{x^2-y^2}$ 
and the $\sqrt{2}\times \sqrt{2}$ d$_{xy}$ pairings, 
whose order parameters are 
$c_{{\bf k}\uparrow}c_{-{\bf k}\downarrow}(\cos k_x-\cos k_y)$ and
$c_{{\bf k}\uparrow}c_{-({\bf k+Q})\downarrow}(\sin k_x\sin k_y)$, 
respectively, with ${\bf Q}\equiv (\pi,\pi)$.  
If they both have long-range orders, we should 
take $(c_{{\bf k}\uparrow}, c_{{\bf k+Q}\uparrow})$ and 
$(c_{-{\bf k}\downarrow}, c_{-({\bf k+Q})\downarrow})$ as basis 
to diagonalize the $2\times 2$ order parameter matrix to have
\[
\Delta_{\pm} ({\bf k})=\pm \sqrt{A(\cos k_x-\cos k_y)^2+B(\sin k_x\sin k_y)^2},
\]
where $A,B>0$.  This form, which is nodeless, 
is similar to the energy spectrum of the chiral spin state
proposed by Wen, Wilczek, and Zee.\cite{WWZ} The order parameter of the chiral
spin state is defined for $\langle c^\dagger c\rangle$, the hopping
amplitude, while we are here talking about $\langle c c\rangle$, 
the pairing amplitude.  
The corresponding superconducting gap coincides with that of the
d$_{x^2-y^2}+ i$d$_{xy}$ pairing,\cite{Rokhsar,Laughlin} 
but we must stress that the present order 
parameter is real and hence does not break the time reversal symmetry
as in d$_{x^2-y^2}+ i$d$_{xy}$. Thus we end up with a {\it fully-gapped, 
time-reversal-symmetric mixture} of d$_{x^2-y^2}$ and d$_{xy}$ pairings.

As seen in Fig.\ref{fig1} (b), the $\sqrt{2}\times \sqrt{2}$ structure 
of the d$_{xy}$ correlation is not observed in the hole-doped case.
In fact, we have considered a wide variety of cases, some of 
\noindent
which will be published elsewhere, and found that the 
$\sqrt{2}\times \sqrt{2}$ structure in the d$_{xy}$ pairing emerges
only when $(\pm \pi/2, \pm \pi/2)$ lies on the Fermi surface.
Then, the difference in the pairing symmetry between the hole-doped 
and electron-doped cases may be not only due to the relative 
position of the VHS against $E_F$, but may also come from the fact that 
$(\pm \pi/2, \pm \pi/2)$ lies very close to the Fermi surface 
in NCCO. 

The relation of $(\pm \pi/2,\pm \pi/2)$ to the pairing 
having a superstructure has also been suggested 
for the $t$-$J$ model by Ogata quite recently\cite{Ogata}.
Using a variational approach to the $t$-$J$ model, 
he showed that the energy of d$_{x^2-y^2}$ pairing state is lowered 
with a full gap when mixed
with NN extended-s pairing having finite momentum of $(\pi,0)$ or $(0,\pi)$, 
if the system is lightly doped, so that $(\pm \pi/2,\pm \pi/2)$ 
is close to the Fermi surface.\cite{comment2} 
The difference with the present d$_{xy}$ superstructure is that 
the pair here has a finite total crystal 
momentum of $(\pi,\pi)$ resulting in a
$\sqrt{2}\times \sqrt{2}$ structure, while Ogata's extended s-pair 
has a momentum $(\pi,0)$ or $(0,\pi)$ with a $2\times 2$ structure.  
As mentioned above, we have so far found extended s-wave correlations
to be suppressed at large distances, but we believe further calculation 
for various values of parameters is necessary to reveal
the relation between the present result for the Hubbard model and
Ogata's result for the $t$-$J$ model.

In summary, we have shown that the 2D Hubbard model possessing
band features experimentally observed in the cuprates can account
for both the d$_{x^2-y^2}$ pairing for hole doping and
a nodeless pairing for electron doping. 
The fact that the present 
result is obtained for rather small values of $U(\sim t)$ suggests that 
large interactions $(U\gg t)$ may not be essential to 
the occurrence of superconductivity, 
although the strength of the interaction will 
certainly dominate the absolute magnitude of the gap or T$_C$.

We would like to thank  M. Ogata and S. Kashiwaya for useful 
discussions and sending us refs.\onlinecite{Ogata} and \onlinecite{Kashiwaya},
respectively, prior to publication.
Numerical calculations were performed at the Supercomputer Center,
Institute for Solid State Physics, University of Tokyo,
and at the Computer Center of the University of Tokyo.
This work was supported in part by Grant-in-Aid for Scientific
Research from the Ministry of Education
of Japan.

%%%%%%%%%%%%%%%%%%%%  References %%%%%%%%%%%%%%%%%%%%%%%%%%%%%%%%

\end{multicols}

\begin{references}
\bibitem{comment} We will use the words HOL's (LUL's) for the 
occupied (unoccupied) states which are within 0.01 to $E_F$ in energy,
although these states are not necessarily exactly degenerate.
\bibitem{comment2} In this sense, the present $\sqrt{2}\times\sqrt{2}$ 
d$_{xy}$ state also might have some relevance to the lightly 
hole doped situation.
\end{references}
\end{document}